\documentclass[10pt,letterpaper]{article}
\usepackage[top=0.85in,left=2.75in,footskip=0.75in]{geometry}

\usepackage{amsmath,amssymb}

\usepackage{changepage}

\usepackage[utf8x]{inputenc}

\usepackage{textcomp,marvosym}

\usepackage{booktabs}
\usepackage{lipsum}
\usepackage{multirow}

\usepackage{cite}

\usepackage{nameref}
\usepackage[hidelinks]{hyperref}
\usepackage[right]{lineno}

\usepackage{microtype}
\DisableLigatures[f]{encoding = *, family = * }

\usepackage[table]{xcolor}
\usepackage{xcolor}
\usepackage{array}

\newcolumntype{+}{!{\vrule width 2pt}}

\newlength\savedwidth

\usepackage[aboveskip=1pt,labelfont=bf,labelsep=period,justification=justified,singlelinecheck=off]{caption}
\renewcommand{\figurename}{Fig}

\makeatletter
\renewcommand{\@biblabel}[1]{\quad#1.}
\makeatother

\usepackage{xcolor}
\usepackage{soul}
\usepackage{xspace}
\definecolor{reddish}{HTML}{FBB4AE}
\definecolor{blueish}{HTML}{B3CDE3}
\definecolor{magentish}{HTML}{FF00AA}
\definecolor{greenish}{HTML}{a1d99b}

\usepackage{lastpage,fancyhdr,graphicx}
\usepackage{epstopdf}
\usepackage[normalem]{ulem}
\fancyheadoffset[L]{2.25in}
\fancyfootoffset[L]{2.25in}
\newcommand{\eg}{{\it e.g.}\xspace}

\begin{document}
\vspace*{0.2in}

\begin{flushleft}
{\Large
\textbf\newline{Vector fields as a framework for modelling the mobility of commodities
}
}
\newline
Sima Farokhnejad\textsuperscript{1,*},
Ang\'elica S. da Mata\textsuperscript{2},
Mariana Macedo\textsuperscript{3,4},\\
Ronaldo Menezes\textsuperscript{1,*}
\\
\bigskip
\textbf{1} Department of Computer Science, University of Exeter, UK\\
\textbf{2} Department of Physics, Federal University of Lavras, Minas Gerais, Brazil\\
\textbf{3} Department of Data Science, Northeastern University, London, UK\\
\textbf{4} Khoury College of Computer Sciences, Northeastern University, Boston, United States \\
\bigskip

$^{*}$ sfarokhnejad@biocomplexlab.org, r.menezes@exeter.ac.uk

\end{flushleft}

\section*{Abstract}

Commodities, including livestock, flow through trade networks across the world, with trajectories that can be effectively captured using mobility pattern modelling approaches similar to those used in human mobility studies. However, documenting these movements comprehensively presents significant challenges; it can be unrealistic, costly, and may conflict with data protection regulations. As a result, mobility datasets typically contain inherent uncertainties due to sparsity and limitations in data collection methods.
Origin-destination (OD) representations offer a powerful framework for modelling movement patterns and are widely adopted in mobility studies. However, these matrices possess inherent structural limitations: locations absent from the OD framework lack spatial information regarding potential mobility directions and intensities. This spatial incompleteness creates analytical gaps across different geographical granularities, constraining our ability to fully characterise movement patterns in underrepresented areas.
In this study, we introduce a vector-field-based method to address these persistent data challenges, transforming OD data into vector fields that capture spatial flow patterns more comprehensively, enabling us to also study mobility directions in a robust manner.
We utilise cattle trade data from Minas Gerais, Brazil, as our case study for commodity flows. This region's extensive livestock trading network makes it an ideal test case for our vector-field methodology. Cattle movements are particularly significant as they directly impact disease transmission, including foot-and-mouth disease. Accurately modelling these flows enables more effective disease surveillance and control strategies, with implications for both animal health and economic stability.
Our vector-field approach reveals fundamental patterns in commodity mobility and can infer movement information for unrepresented locations; a critical factor in modelling scenarios such as disease spread. Our approach offers an alternative to traditional network-based models, enhancing our capacity to infer mobility patterns from incomplete datasets and advancing our understanding of large-scale commodity trades.

\section*{Introduction}

 Modelling mobility patterns has long been the focus of many studies~\cite{barbosa2018human,chang2021mobility,verginer2020cities,Wang2019urbanmobility,soriano2022modeling}, with network-based methods playing a central role in capturing the flow of commodities across the world. While networks have proven useful in many contexts, they often miss key nuances, particularly when applied to large and complex datasets. Mobility datasets of commodities, such as those tracking livestock trade, tend to be vast, containing many individual records~\cite{bajardi2012optimizing}, which makes network generation both computationally expensive and difficult to interpret~\cite{lentz2016disease}. These datasets are also subject to uncertainty~\cite{Farokhnejad:Data:2022, zheng2015trajectory}, not only because of errors in data collection but also because it is practically impossible to capture every detail of large-scale movements completely. This uncertainty is inherent to the nature of mobility data~\cite{gonzalez2008understanding}, meaning that complete accuracy in tracking movement is unattainable. For instance, most work in human mobility rely on specific datasets that often capture only a fraction of reality, such as call detail records (CDRs), extended detail records (XDRs), and location-based social networks (LBSNs)~\cite{wang2011human,blondel2015survey}. As a result, modelling with incomplete data -- often the only available option -- can introduce inaccuracies and uncertainties in contexts such as mobility analysis or disease spread prediction. This highlights the importance of addressing the issue of missing data and exploring methods to enhance the accuracy of mobility modelling and its applications.

Another common limitation is that network models typically rely on origin-destination (OD) matrices, which focus on the starting and ending points of movements~\cite{mamei2019evaluating}. However, this representation is also incomplete by design. Locations that are not explicitly involved in the recorded trades are ignored, even though they may play significant roles in the overall flow of commodities. For example, intermediate locations through which commodities are transported can still influence the dynamics of the trade, yet they are not represented in OD matrices. This leads to gaps in our understanding of movement patterns, particularly in regions where data are missing or uncertain. If these unknown locations are cities or towns, for instance, we are unable to assess their importance in the holistic view of the system.

\begin{figure}[ht]
\centering
\includegraphics[width=\textwidth]{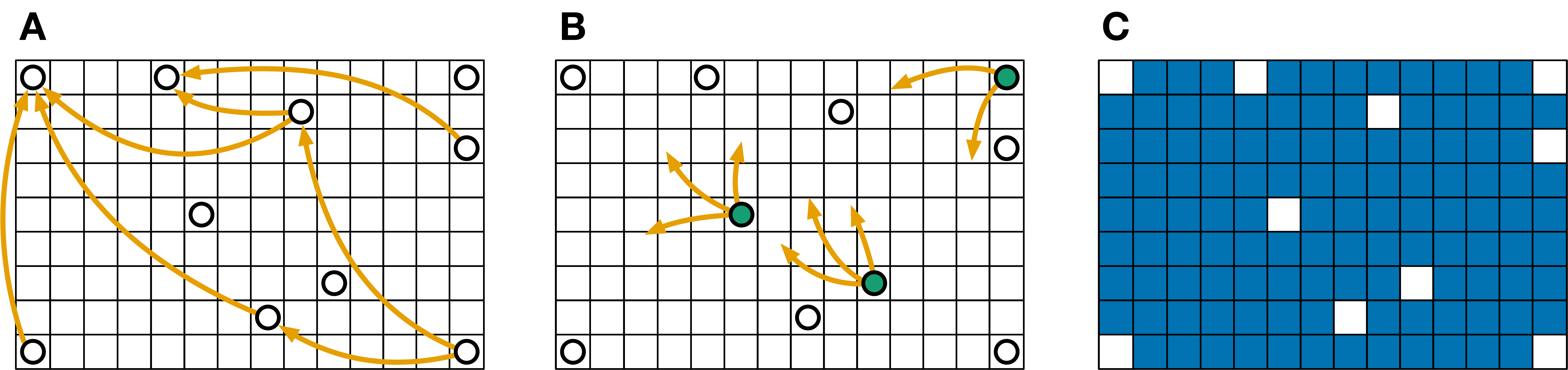}
\caption{{\bf Data sparsity can impact network modelling as it can disregard important locations.} {\bf (A)} Typical OD model of commodity movements where locations not included in the recorded movements are excluded entirely from the network. {\bf (B)} Certain areas that serve as spatial junctions are left out simply because they are neither origins nor destinations, despite their potential importance in facilitating flow. {\bf (C)} Data gaps result in missing regions (\eg, blue cells), limiting our ability to assess their significance within the overall movement system. This toy example illustrates how OD models may fail to capture the full spatial complexity of commodity flows.}  
\label{fig:netVflow}
\end{figure}

Take, for example, the scenario in \figurename~\ref{fig:netVflow}. The map in panel A represents the typical origin-destination (OD) model of commodity movements, where only the start and end points of movements are captured. In this representation, certain locations that may serve as critical intersections for multiple movements are excluded because they are neither origins nor destinations (illustrated as green nodes in \figurename~\ref{fig:netVflow}B). Even though these locations are not directly involved in the movements, they may play an essential role in the overall flow of commodities. In our modelling, we consider the movements as a series of vectors and calculate the resulting vector that represents that region. Once that is done, we can use interpolation to calculate all the missing vectors for regions that do not have information (filling the blue cells in \figurename~\ref{fig:netVflow}C). The interpolation approach is based on the reasonable assumption, particularly in real-world scenarios, that neighbouring locations are likely to participate in and influence similar activities. Thus, our approach aims to address these data gaps by inferring the likelihood of flow for unrepresented locations, offering a more comprehensive view of the system dynamics. Furthermore, interpolation accuracy and robustness depends on the amount of missing information as explained later in \figurename~\ref{fig:vector_flow_robustness}. 

We propose a vector field-based method as an alternative to traditional network models for analysing commodity flows/mobility. By transforming OD data into vector fields, our approach captures likely movement directions even in unrepresented areas, offering a more realistic view of spatial dynamics. This is particularly valuable because, in practice, it means that if future trade nodes are established in these previously unrepresented locations, we would already have a preliminary indication of the likely directions their commodities will flow towards, providing a foundation for further analysis.

The application of vector fields to commodity flows is novel, constituting a significant methodological advance in the study of commodity movement. However, the concept of vector fields itself is not new~\cite{liu2005patterns}. Vector fields have been utilised in disciplines such as meteorology, where they are used to represent wind direction and velocity across geographical regions~\cite{wittenbrink1996glyphs}. These fields allow meteorologists to visualise and analyse weather patterns, predict storm paths, and model airflow. By translating commodity trade movements into vector fields (which we refer to as commodity flow maps in this paper), we can build on the extensive foundation of methods developed for weather mapping and other fields that deal with spatial flows~\cite{andrienko2013visual}. Established techniques such as vector interpolation, field integration, and streamline generation---routinely used in weather analysis---can be directly adapted to the study of commodity flows, enhancing both the accuracy and efficiency of our approach. This cross-disciplinary application enables a more dynamic and flexible analysis of trade data, where unrepresented areas can still contribute to the overall understanding of mobility patterns.

To demonstrate the utility of our approach, we apply this vector field methodology to cattle trade data from the Brazilian state of Minas Gerais. Minas Gerais is one of the largest cattle-trading regions in Brazil, with a complex and extensive trade network. The scale of this trade activity, coupled with the incomplete nature of available data, makes it an ideal test case for evaluating the effectiveness of our methodology in converting OD data into vector fields that represent commodity flows.

One key direct application of this method is in understanding the spread of diseases, such as those affecting livestock like foot-and-mouth disease \cite{tomley2009livestock,volkova2010potential,Ferguson2001}. In traditional models, the lack of data on certain locations limits our ability to predict how a disease might spread through a location that is not originally represented as origin or destination. Vector fields, however, allow us to estimate potential commodity (livestock) flows through these unrepresented areas, offering a more comprehensive tool for assessing risk. This can be crucial for decision-makers who need to predict and manage the spread of diseases across regions with incomplete data.

To analyse the flow of commodities over time, we rely on two essential parameters: entropy~\cite{shannon1948mathematical} and cosine similarity~\cite{beggs2012being}. In this paper, entropy quantifies the diversity of trade directions throughout the year, helping to distinguish between regions with stable and fluctuating trade behaviours. Meanwhile, cosine similarity captures the temporal evolution of commodity flow directions across different timescales---monthly, weekly, seasonal, and yearly---allowing us to identify patterns of consistency or variability in trade routes. These measures provide a robust framework for understanding the broader structure of commodity flows and their temporal shifts within our vector field representation.

Beyond these statistical analyses, the vector field approach offers a comprehensive representation of commodity flows at multiple levels of granularity, enabling diverse experimental applications. A key aspect of our analysis involves identifying critical points within the vector field~\cite{smolik2016vector}, which serve as focal points for understanding the local and global structure of commodity movements. By studying their behaviour, we gain insights into trade dynamics that are not captured by traditional network models. The classification of these critical points provides a structured understanding of the field's topology, and by assuming smooth transitions between regions, we can construct a simplified yet informative model of the overall commodity flow system.

Our results show that vector fields offer an efficient and adaptable framework for analysing large-scale commodity mobility datasets. This is demonstrated through our successful interpolation of flow patterns in unrepresented areas and the identification of critical points that reveal underlying structural patterns in the cattle trade network of Minas Gerais. By focusing on spatial flows rather than discrete network nodes and edges, vector fields provide a more flexible approach to capture movement patterns. This method is particularly relevant in cases where uncertainty and missing data are unavoidable, offering a valuable alternative to traditional network-based models. 

\section*{Materials and methods}

\subsection*{Data formatting}

Our approach to generating vector fields uses origin–destination (OD) data as input, which is the most common publicly available mobility data format. This data can be defined at different time windows and spatial resolutions. By segmenting the data into appropriate temporal intervals and spatial cells, we can project movements onto a grid and subsequently generate vectors that represent aggregate commodity flows between these defined areas. This process is not limited to any single commodity but is applicable to any scenario where movement data can be represented as OD pairs, making it a versatile tool for studying the flow of commodities, goods, or even people.

This general method is particularly valuable when dealing with imperfect or incomplete information. In many practical situations, data gaps and uncertainties exist. That is, certain locations might lack explicit records, or the data may not cover every possible route of movement. The vector field approach addresses these challenges by inferring likely commodity flows and filling spatial gaps using interpolation techniques. As a result, even with partial or uncertain OD data, this methodology offers a comprehensive view of movement patterns, aiding in the analysis and decision-making processes across diverse applications.

The~\nameref{app:Appendix} provides a detailed description of how our vector field approach is applied to the movement of cattle as a commodity. While our methodology is general and applicable to any OD-defined movement data, we later focus on cattle trade in the \nameref{sec:results} section to illustrate the power and versatility of the approach in a concrete scenario.

\subsection*{Creating vector fields}

\begin{figure}[ht]
\centering
\includegraphics[width=.8\textwidth]{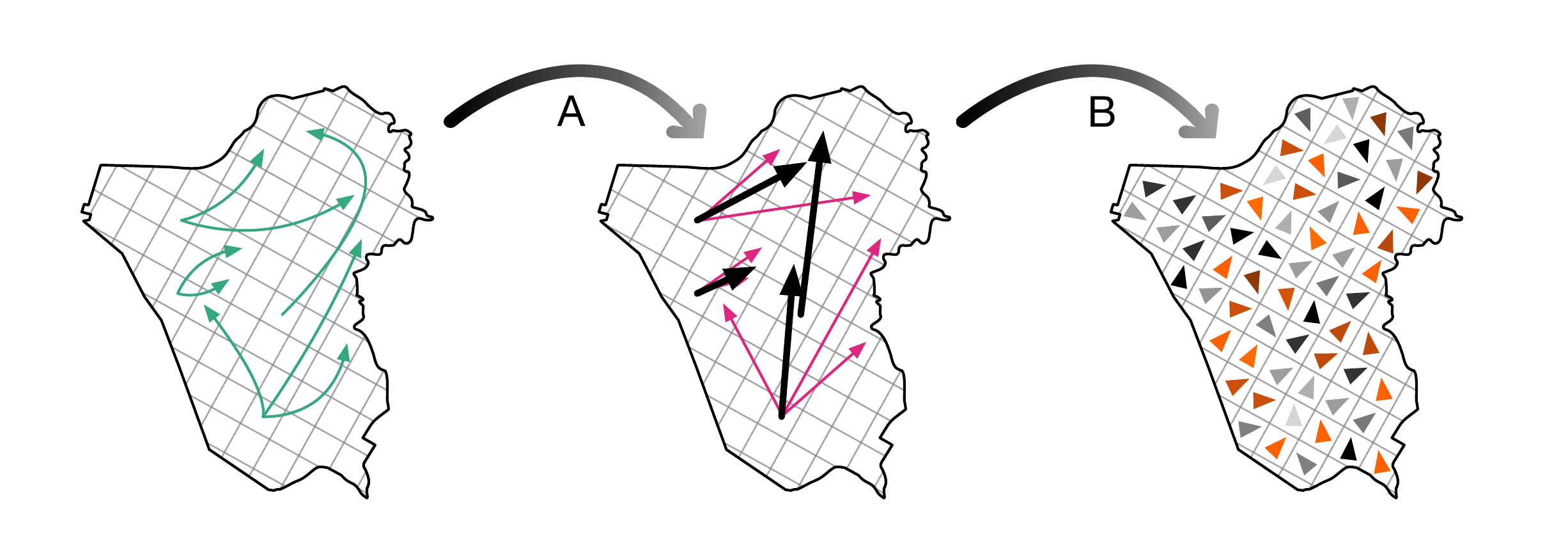}
\caption{\textbf{Generating a vector field from origin-destination data.} Given locations where we have information (origins/destinations of commodity movements represented by directed arrows) shown in the map on the left, \textbf{(A)} we can interpret each origin-destination pair as a vector and calculate the resulting vector of all movements from each particular location. Once this is done for all locations with available data, \textbf{(B)} we can interpolate the missing information, illustrated in the rightmost figure as example arrows shown in shades of orange. The final result is a complete vector field where all locations (cities, for instance) have a vector with direction and magnitude. In this example, the arrows on the rightmost map represent direction and the shade represents the magnitude of the vector.}
\label{fig:flow_network}
\end{figure}

We transform an origin-destination network into a vector field for the region of interest based on the following steps. The first step is to choose suitable temporal and spatial granularities, specifying both the timeframe for trades and how the region is partitioned. We then project the network (i.e. trades within the chosen timeframe) onto this spatial scheme so that each node lies within a particular cell (the first map on the left in \figurename~\ref{fig:flow_network}). For each cell, all outgoing edges are treated as vectors (shown in pink in the middle map in \figurename~\ref{fig:flow_network}), which are combined into a single resultant vector (shown in black in the middle map in \figurename~\ref{fig:flow_network}) whether through averaging, summation, or other methods. Repeating this process for every cell with outgoing edges produces a partial vector field. As some cells do not generate outgoing edges, we employ interpolation to fill in the missing vectors and complete the full vector field (shown in shades of orange in the rightmost map in \figurename~\ref{fig:flow_network}).

To assess the global behaviour of vector field patterns, it is necessary to interpolate values for locations without originating edges, thereby completing the field. Interpolation constructs vectors at new points based on known (or measured) vectors, enabling a continuous representation of commodity flows across the entire region. We opt for a triangle-based interpolation owing to its simplicity and suitability for our sparse networks\cite{watson1984triangle, watson1985refinement}. This method estimates vector values in grid cells lacking \textit{a priori} data, ultimately yielding a complete vector field (right map in \figurename~\ref{fig:flow_network}). For technical details on the interpolation method, see the~\nameref{app:Appendix}.

Vector field analysis offers two complementary approaches for understanding commodity mobility patterns: global flow exploration and critical point identification.

Global exploration of flow patterns focuses on how a region's mobility behaviour changes through its dominant flow vector, as we later demonstrate with cattle trade in Section~\nameref{sec: Diversity and Regularity of Cattle Flows}. This approach examines the vector computed for each region, following the method outlined in \figurename~\ref{fig:flow_network}. Changes in the dominant direction of a region's vector, as well as similarities between its vector dynamics and those of other regions, can be effectively observed through vector field representations. Temporal patterns can be quantified using vector directions by applying measures such as cosine similarity and entropy (see the~\nameref{app:Appendix} for the methodological details), while spatial correlations based on vector magnitudes can highlight groups of cities with comparable trade behaviour, including clusters that reflect different trading distances—such as short-, medium-, and long-distance commodity flows. Examining these spatial patterns can identify areas with consistently high or low trade intensity. Comparing these characteristics across multiple time intervals allows clustering regions with similar dynamic behaviours, and identifying key roles in overall mobility patterns.

Critical points in vector fields, where the flow vanishes, reveal essential structural characteristics of mobility patterns. These points can be classified as attracting or repelling points based on the eigenvalues of the Jacobian matrix of the field~\cite{smolik2016vector}. We illustrate the presence of critical points within a vector field, highlighting regions of flow convergence (see Fig.~A~\ref{fig:critical_points_toy}B in \nameref{app:Appendix}). These points are explored in our analysis of cattle trade (see Section \nameref{sec:criticsl_points}). However, their full interpretation in the context of commodity flows remains a question for future research.

In summary, transforming networks into vector fields offers a novel way to model the dynamics of commodity flows. In the context of cattle trade (see Section~\nameref{sec:results}), this transformation helps reveal seasonal patterns, evolving critical points, and changes in flow dynamics over time.

\section*{Results}
\label{sec:results}

We apply our proposed vector field approach to cattle trade data to demonstrate how this method reveals complex mobility patterns. By transforming raw trade data into geographic vectors, we identify key insights such as dominant flow directions, trade hubs, and critical points. These insights are often missed by traditional network models.

Our analysis spans municipality and micro-region levels, using techniques like entropy, cosine similarity, and spatial autocorrelation (see~\nameref{app:Appendix} for further details). We then showcase the method's ability to handle incomplete data, uncover commodity movement patterns, and support applications such as planning, risk assessment, and disease control.

\subsection*{Robustness of vector fields}

\begin{figure}[!ht]
\centering
\includegraphics[width=.8\textwidth]{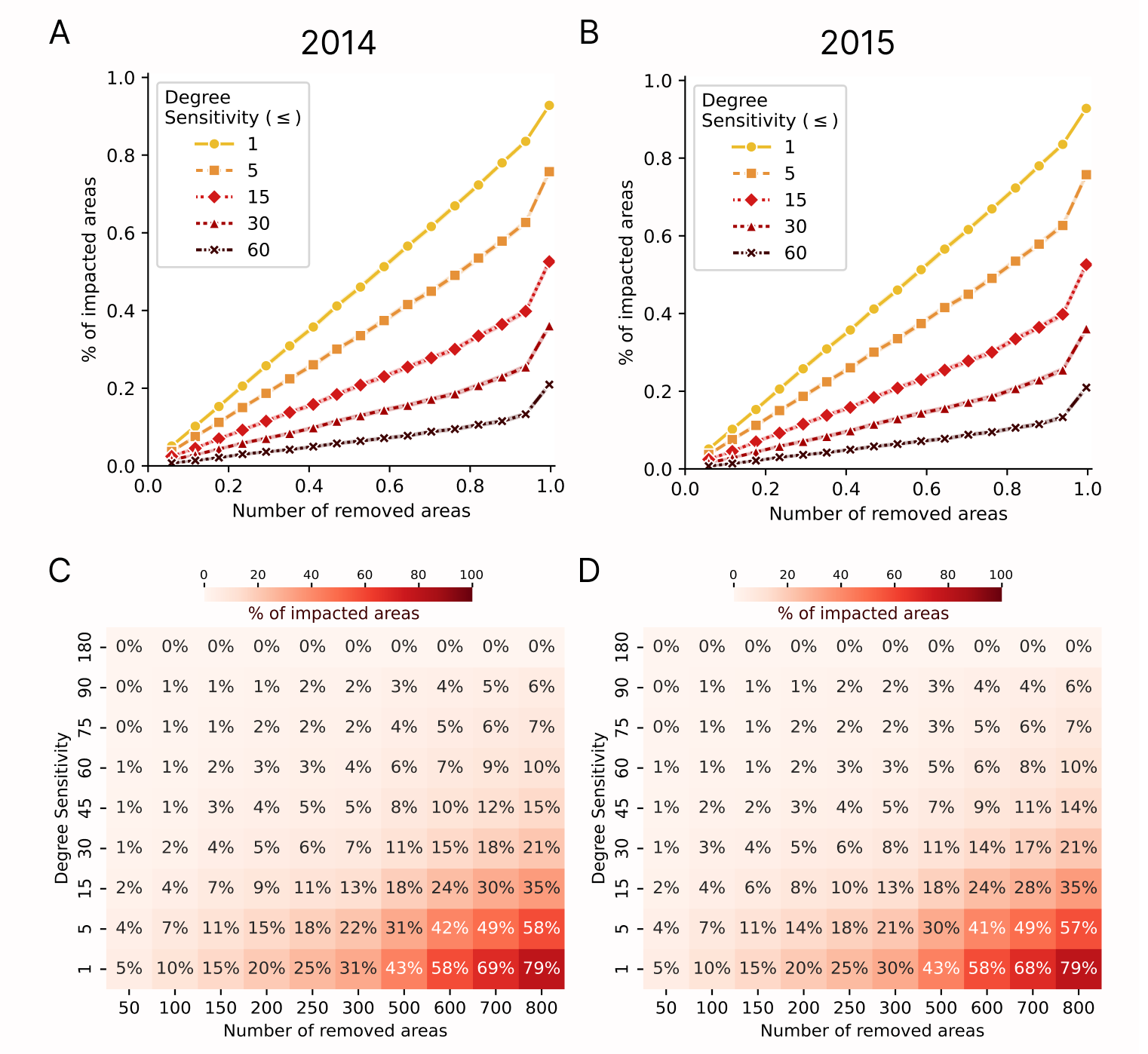}
\caption{\textbf{Robustness of vector field estimations under spatial data removal.} Proportion of regions exhibiting directional changes as a function of the number of randomly removed areas for 2014 (A, C) and 2015 (B, D). The results demonstrate the robustness of vector field estimations despite spatial data gaps, highlighting the reliability and efficiency of our approach across different years.}
\label{fig:vector_flow_robustness}
\end{figure}

Before analysing the trade data, we assess the robustness of our vector field reconstruction approach by systematically removing spatial regions and evaluating the extent to which original flow directions are preserved. This procedure simulates the presence of missing trade data and demonstrates the capacity of vector fields to interpolate commodity flows in unobserved areas. As shown in \figurename~\ref{fig:vector_flow_robustness}, we quantify robustness in terms of the proportion of vectors affected and the angular deviation in degrees. Three key findings emerge: \textit{(i)} robustness remains consistent across years (fluctuations tend to be less than 3\%), \textit{(ii)} most deviations are below 15 degrees, and \textit{(iii)} over 60\% of the spatial information must be removed before more than half of the vectors experience any change. These findings indicate that our approach is highly resilient to spatial sparsity, with typical deviations constrained to minimal angular shifts.

\subsection*{Cattle commodity flows and fields}
\label{sec-Cattle Flows and Fields}

We use the cattle trade data at two levels of granularity: municipalities and micro-regions. For the municipality level (Fig.~A~\ref{fig:vector_map_city_mic}A in the~\nameref{app:Appendix}), we begin by isolating a single municipality (shown in grey in Fig.~A~\ref{fig:vector_map_city_mic}A in the~\nameref{app:Appendix}) and considering all trades originating from it to other municipalities. We do not consider local trades that occur within the same municipality (self-loops). Each trade is transformed into a vector starting from the centre of a municipality and pointing towards the centre of the destination municipality. By averaging (or summing) these vectors, we obtain a single representative vector for the municipality, resulting in a partial vector field that captures the predominant flow direction and intensity for that location over a given time period. This process exemplifies how scattered trade data are aggregated into a coherent vector field at the municipal scale.

To move from this partial representation to a comprehensive view, we apply triangle-based interpolation as it preserves the spatial relationship (structure) and avoids artefacts by considering local patterns~\cite{watson1984triangle,watson1985refinement} (further details in the~\nameref{app:Appendix}). This interpolation fills in gaps where direct trade data are absent, creating a complete vector field over the entire municipal region that faithfully represents the continuous flow patterns of cattle movements.

\subsection*{Diversity and regularity of (cattle) commodity flows}
\label{sec: Diversity and Regularity of Cattle Flows}

We analyse the monthly vector fields from the trade data for two distinct spatial divisions: municipalities and micro-regions. Our goal is to assess how the directions of these commodity flows evolve over time and across different regions. To achieve this, we use two key metrics:
\begin{description}
    \item [Entropy:] measures the diversity of trade directions over the year. In particular, we use Shannon entropy, which quantifies how often trades occur across different directions (see the~\nameref{app:Appendix} for the details of Shannon entropy). 
    \item [Cosine similarity:] captures the similarity (or dissimilarity) in commodity flow direction between pairs of consecutive months. A value near \(1\) indicates that two monthly vectors point in nearly the same direction (high regularity), while a value near \(-1\) indicates they point in opposite directions (high variability). 
\end{description}

During each year, every municipality (or micro-region) has 12 monthly vectors of trade flows. We categorize each vector into one of four quadrants based on its direction in the Cartesian plane, thus reducing the original set of 12 vectors to a sequence of 12 quadrant values. We then compute the Shannon entropy of this set, capturing the diversity or unpredictability of trade directions over the year.

\begin{figure}[ht]
\centering
\includegraphics[width=\textwidth]{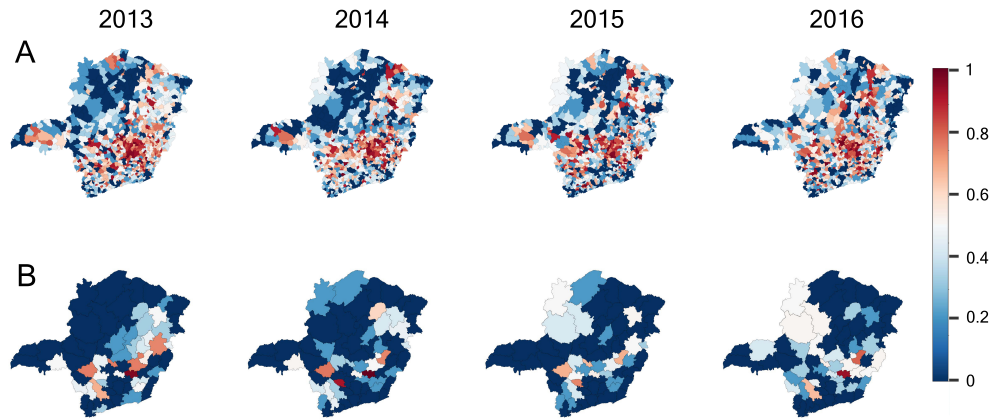}
\caption{\textbf{Entropy of monthly vectors for (A) municipalities and (B) micro-regions.} 
Each location's 12 monthly vectors are grouped into four directional clusters. Shannon entropy then yields a single value per municipality or micro-region, which is colour-coded on the map to reflect diversity in commodity flow directions over the year. The entropy is normalised for visualisation purposes using MinMax normalisation.}
\label{fig:entropy-city}
\end{figure}

\figurename~\ref{fig:entropy-city} presents entropy heatmaps of municipalities and micro-regions. Many municipalities (and micro-regions) in the North exhibit lower entropy (darker blue shades, closer to zero) indicating that trading directions tend to be more stable. Some regions with higher entropy (darker red shades) have more frequent shifts in commodity\footnote{Note, the term 'commodity' refers to 'cattle as commodity' in the results section given that cattle trade has been used as the case study.} flow direction, indicating greater unpredictability. As the highest inter-regional trade volume in Minas Gerais occurs between the North and West regions, the unpredictability is mostly observed in the South-east—regions with less established trading relationships.

\begin{figure}[!ht]
\centering
\includegraphics[width=1\textwidth]{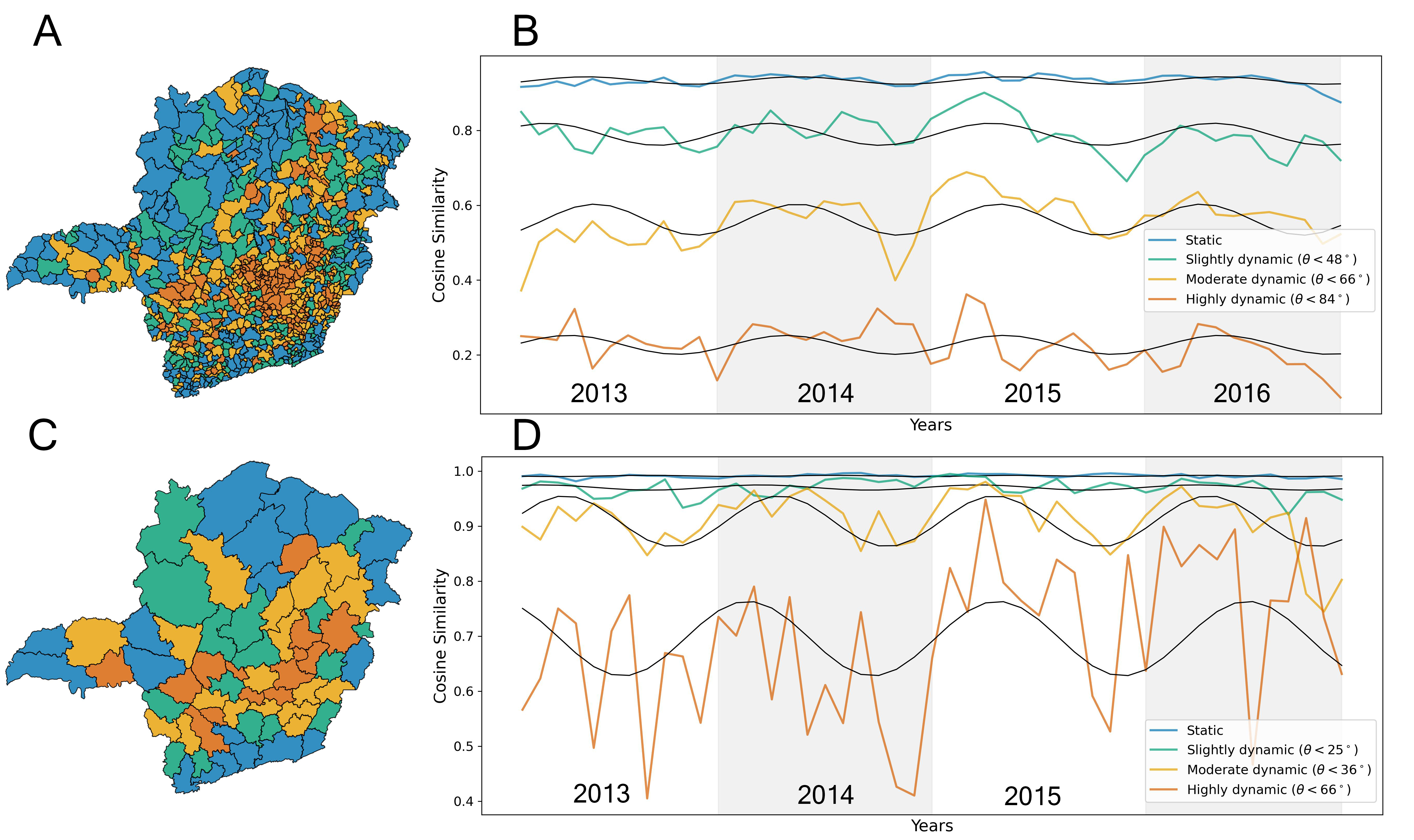}
\caption{{\bf Cosine-based clustering of (A, B) municipalities and (C, D) micro-regions.} Using K-medoids clustering (k=4), locations are grouped by their cosine similarity values. Clusters, indicated by distinct colours, reveal that in many peripheral locations the commodity flow directions remain stable over the years.}
\label{fig:cs_cluster_city_mic}
\end{figure}

To assess how trade directions evolve over short time intervals, we use the cosine similarity between pairs of consecutive monthly vectors for each municipality or micro-region (e.g., from January 2013–February 2013 to November 2016–December 2016). This method allows us to focus on the structure and direction of commodity flows rather than their magnitude, making it well-suited for detecting shifts or redirections in trade patterns over time. We then use these cosine similarity values to cluster municipalities and micro-regions based on their month-to-month flow direction patterns (\figurename~\ref{fig:cs_cluster_city_mic}). We identify four well-established clusters: from highly static (shades of blue) to highly dynamic (shades of orange). Different colours in the cluster maps indicate groups of regions that share similar directional dynamics over time. In essence, we group areas (cities or microregions) whose flow directions not only change month-to-month by similar angles, but also follow similar patterns of change throughout the year. This captures both the magnitude and rhythm of directional shifts. \figurename~\ref{fig:cs_cluster_city_mic} displays the average cosine similarity values of municipalities or micro-regions within each cluster. Detailed cosine similarity values for individual areas are provided in our previous study~\cite{farokhnejad2023using}.

In both municipalities (\figurename~\ref{fig:cs_cluster_city_mic}A-B) and micro-regions (\figurename~\ref{fig:cs_cluster_city_mic}C-D), a large dominant cluster emerges, consisting of areas with relatively stable flow directions (cosine values close to 1), indicating regular and predictable cattle movement. However, there are cities and micro-regions where flow directions vary significantly---more than 45 degrees across several monthly intervals over the four-year period. Even in these more dynamic cases, the angular changes exhibit a periodic, sinusoidal-like pattern, with a noticeable drop in cosine similarity---indicating increased angular change—towards the end of each year (e.g., November–December), as shown in \figurename~\ref{fig:cs_cluster_city_mic}B and~\ref{fig:cs_cluster_city_mic}D.

This observed rhythmicity reveals a deeper temporal structure in trade patterns, going beyond static flows or random fluctuations. Even if the peak of directional change shifts in time across regions (e.g., occurring earlier or later in the year), the underlying wave-like structure remains, highlighting the existence of region-specific temporal rhythms. These findings indicate a higher level of predictability in cattle movement dynamics and demonstrate that the combined use of entropy (to measure directional diversity) and cosine similarity (to assess regularity) can be effectively extended to analyse specific types of cattle trade transactions.

\subsection*{Going beyond direction in vector fields}

In our previous findings, we focused primarily on the direction of vector fields. While this provided valuable insights, it overlooked a key dimension: magnitude. Two vectors may share the same direction yet represent very different dynamics---one indicating a short trade distance, the other a much longer one. By incorporating vector magnitudes into our analysis, we expand the study to capture both spatial and temporal variations in trade distances, offering a more comprehensive view of mobility patterns. Specifically, we investigate whether regions tend to have trade distances similar to those of their neighbouring areas. This spatial analysis sheds new light on the role of geographic proximity in shaping trade behaviour and commodity movement patterns.

\begin{figure}[!ht]
\centering
\includegraphics[width=.7\textwidth]{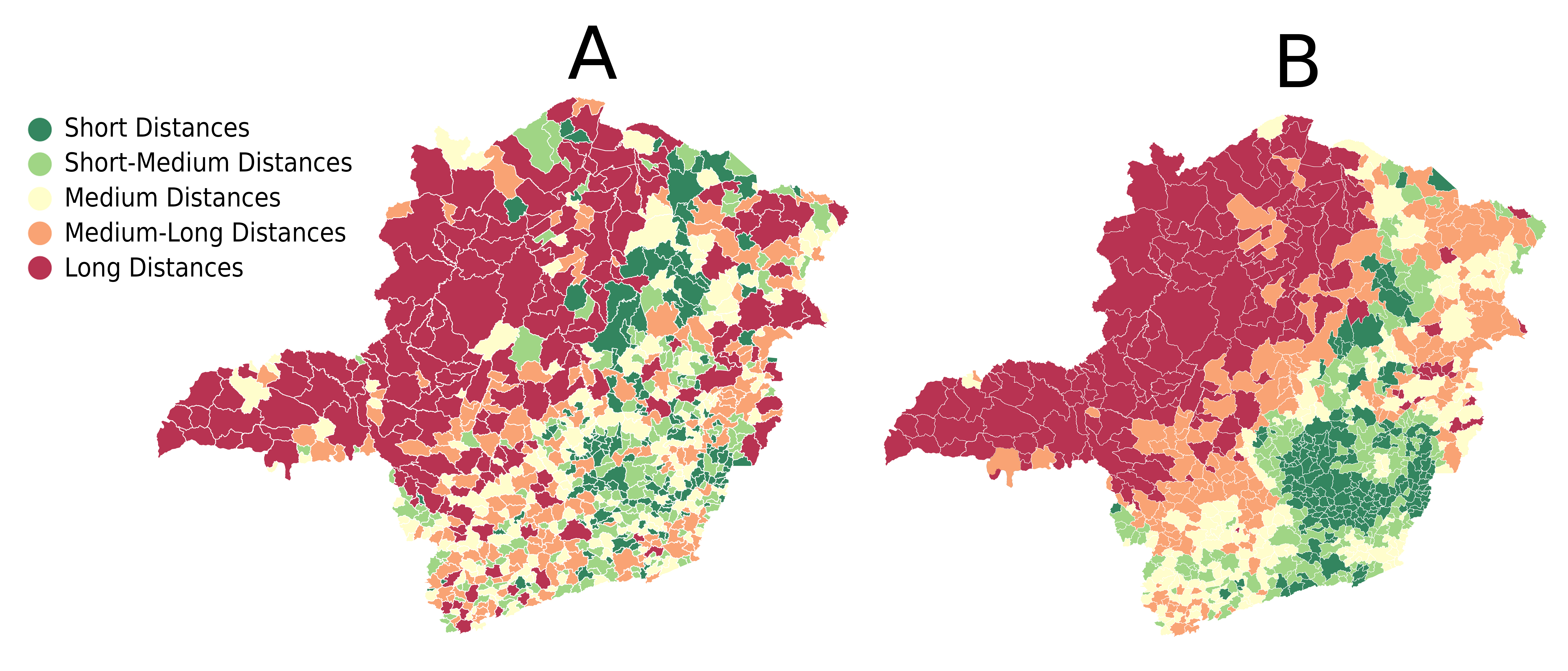}
\caption{\textbf{Maps of trading distances and spatial lag values.} 
\textbf{(A)} Each municipality's trade distance is represented by summing its outgoing vectors, categorised into five distance clusters with distinct colours. 
\textbf{(B)} Spatial lag values, calculated with Queen-based spatial weights, highlight municipalities that differ markedly from their neighbours, revealing `doughnut' patterns, where low values are encircled by higher ones, and `diamond' patterns, where high values are encircled by lower ones. These maps are generated using the flow data from January 2013.}
\label{fig:vector_size_spatial_lag}
\end{figure}

\figurename~\ref{fig:vector_size_spatial_lag} illustrates the spatial variation in vector magnitudes across municipalities and the corresponding spatial lag values. These maps highlight municipalities with significantly higher or lower vector magnitudes than their neighbours, identifying notable patterns such as `doughnuts' (low values surrounded by high values) and `diamonds' (high values surrounded by low values). These features are particularly evident in the spatial lag map (\figurename~\ref{fig:vector_size_spatial_lag}B), which provides a more distinct representation of clusters compared to the raw vector magnitude map (\figurename~\ref{fig:vector_size_spatial_lag}A).

\begin{figure}[ht]
\centering
\includegraphics[width=.9\textwidth]{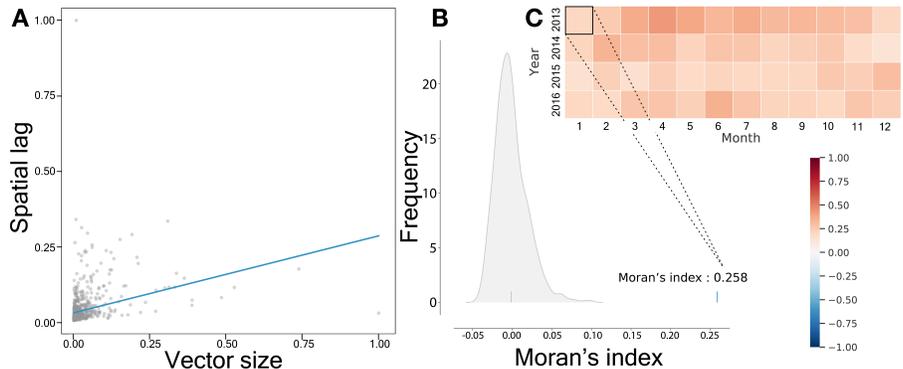}\\[5pt]
\caption{{\bf Global Moran's $I$ calculation.} 
\textbf{(A)} Scatter plot of vector magnitudes against spatial lag with a fitted line. 
\textbf{(B)} To evaluate the significance of the observed pattern, we compared the Moran's $I$ value calculated from the vector field to those from 1000 simulations in which vector magnitudes were randomly shuffled among municipalities. The observed Moran's $I$ value of 0.258 was significantly higher than the simulated values, indicating a non-random spatial structure. \textbf{(C)} The heatmap shows Moran's $I$ values for monthly vector fields across different years. Since no blue shades appear in the heatmap, it indicates that all calculated Moran's $I$ values are positive, suggesting positive spatial autocorrelation---meaning that municipalities with high vector magnitudes tend to be located near each other.}
\label{fig:Moran_I}
\end{figure}

To quantify spatial relationships, we compute Moran's I, a global measure of spatial autocorrelation. The Moran scatter plot (\figurename~\ref{fig:Moran_I}A) reveals a positive correlation between vector magnitudes and their spatial lags, indicating a tendency for similar values to cluster spatially. The observed Moran's I value (\figurename~\ref{fig:Moran_I}B) is statistically significant, suggesting that vector magnitudes are not randomly distributed—also visually evident in \figurename~\ref{fig:vector_size_spatial_lag}. This spatial clustering persists over time, as shown in the Moran's I heatmap (\figurename~\ref{fig:Moran_I}C), which highlights consistent autocorrelation across monthly vector fields.

By analysing vector magnitudes alongside directions, we reveal deeper patterns in trade dynamics and regional differences. While demonstrated here with cattle trade, this vector field approach is broadly applicable to other commodity flows, offering valuable insights for logistics planning and policy development.

\subsection*{Critical points in cattle vector field}
\label{sec:criticsl_points}

To uncover dynamic movement patterns, we use interpolation methods to estimate vectors between known points, transforming scattered data into continuous vector fields. These fields not only capture the flow of commodity trade but also reveal regions of attraction and repulsion, offering valuable insights into spatial dynamics---particularly in understanding the spread and containment of epidemics. By analysing our vector field data, we identify areas where trade concentrates (e.g., high-risk zones for epidemics) and where it diverges (e.g., key suppliers or hubs that can facilitate disease spread).

\begin{figure}[ht]
\centering
\includegraphics[width=.9\textwidth]{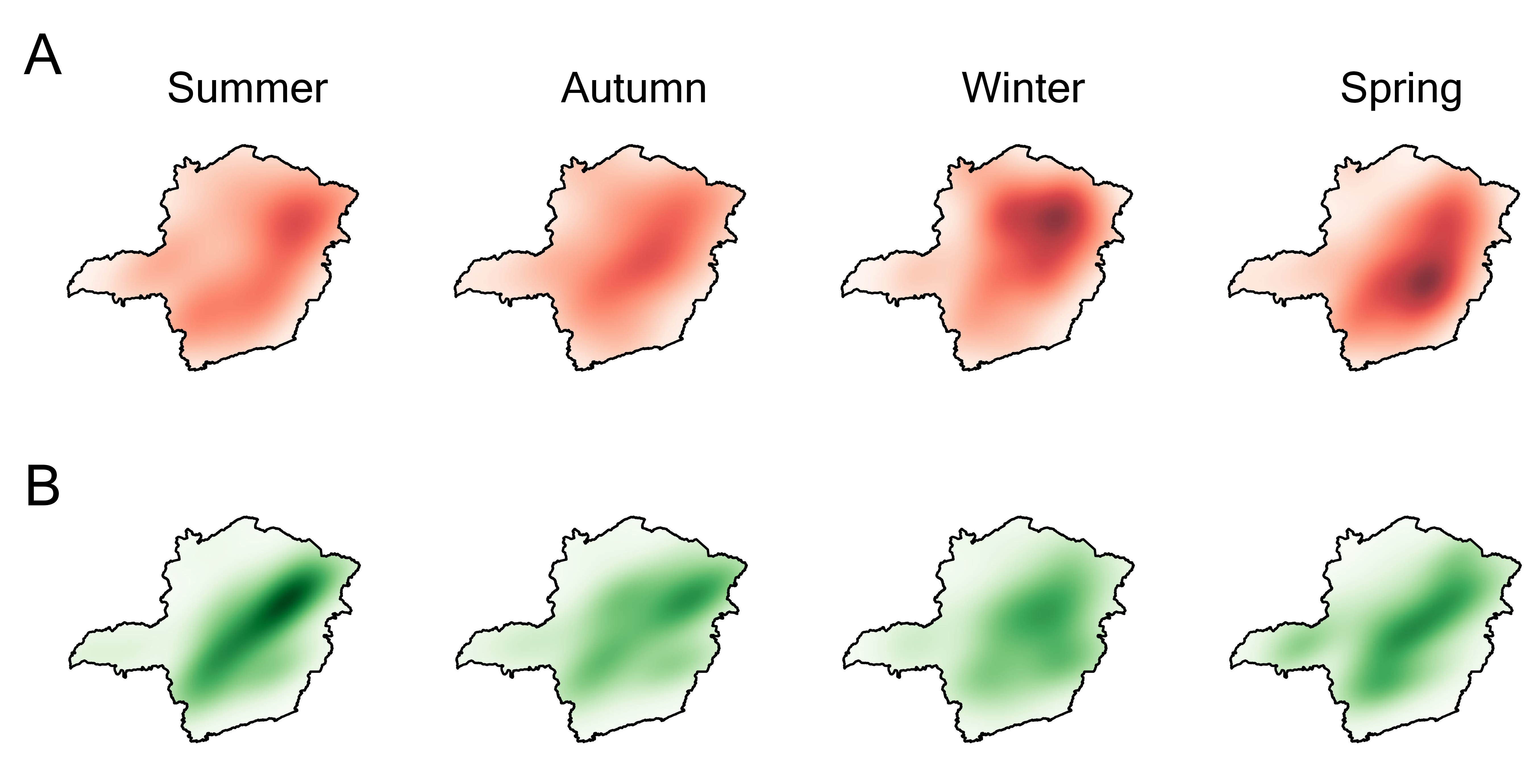}\\[5pt]
\caption{\textbf{Sinks (A) and sources (B) in the cattle trade vector field.} Critical points are shown in the vector field for four seasonal periods, aggregated over a four-year span. The initial vector fields are derived from cattle trades occurring in each season. A triangle-based interpolation method is used to generate a continuous vector field from these initial vectors. Critical points within the field are identified, distinguishing between sinks and sources. These heatmaps indicate the density of sink and source points, with darker shades representing higher concentrations---darker red for sinks and darker green for sources.}
\label{fig:sink_source}
\end{figure}

To quantify these patterns, we calculate the critical points in our seasonal vector fields over a four-year period, distinguishing attracting points (sinks, shown in \figurename~\ref{fig:sink_source}A) from repelling points (sources, shown in \figurename~\ref{fig:sink_source}B). The locations of these sinks and sources shift across seasons, suggesting that while some regions maintain directional stability over time (\figurename~\ref{fig:cs_cluster_city_mic}), disease outbreak management requires careful consideration of the specific month, year, and trade dynamics as these critical points evolve throughout the period.

The movement of cattle follows a structured network, where seasonal shifts and geographical factors play a role in shaping the distribution of sources and sinks. This variability is shaped by a combination of factors, including the geography of the region, the type of premises (e.g., farms, slaughterhouses, or rearing centres), the purpose of the trades, and the broader seasonal trends in cattle trade. For instance, source areas become more prominent during breeding seasons such as spring and summer~\cite{paulino2014brazilian}, while a high density of sink areas corresponds to periods of higher cattle slaughter, such as the months leading up to the New Year. In addition, we observe that regions with a high density of slaughterhouses (reported for 2016~\cite{vale2022regional}) overlap with high-density sink areas in our vector field data during spring.

These patterns highlight not only the mobility within the trade system but also the changing role of different regions in the commodity flow network. Over time, the distribution and strength of sinks and sources may be tied to evolving logistical strategies, shifting market demands, and the structural role of specific regions. The central and western regions frequently act as sources, likely due to their role as intermediate hubs in long-distance transportation. The seasonal vector fields represent cattle movements rather than fixed farm locations. Given the considerable distance between the northern and southern parts of the state, cattle are typically routed through these central source regions before reaching their final destinations. A comparison of the maps in \figurename~\ref{fig:sink_source} and \figurename~\ref{fig:vector_size_spatial_lag}B reveals that sink areas often coincide with short-distance transport zones, while medium- to long-distance transport regions are typically associated with source areas.

\section*{Discussion}
\label{sec:discussion}

This study introduces a vector-field-based framework for analysing commodity mobility, which bridges gaps in traditional origin-destination (OD) approaches by incorporating unrepresented locations into a comprehensive view of mobility patterns. Using cattle trade in the state of Minas Gerais, Brazil, as a case study, we demonstrate that vector fields are a powerful tool for revealing dynamic trade patterns, including: \textit{(i)} clusters of stable trade direction over time, \textit{(ii)} spatial autocorrelation in commodity flow magnitudes, and \textit{(iii)} key attractors and repellers within the region.

A key contribution of this work is the development of robust interpolation techniques that address incomplete and sparse datasets, enabling continuous mobility analysis across spatial regions. Even when more than 50\% of the data is removed, the deviation in vector directions remains below 15 degrees, demonstrating the method's resilience and robustness to sparsity. Additionally, the introduction of entropy and cosine similarity analyses derived from vector fields offers novel ways to measure trade diversity, predictability, and month-to-month flow dynamics. Entropy quantifies the variation in trade directions, identifying regions with stable or unpredictable patterns, while cosine similarity captures temporal changes in flow dynamics, enabling the clustering of regions with similar trade behaviours. Together, these analytical tools enhance the understanding of spatial and temporal dimensions of commodity flows.

Another important contribution of this work is its compatibility with established vector field theories, which open avenues for advanced analyses. Techniques such as critical point dynamics analysis could be explored to understand the temporal evolution of sinks and sources, or streamline visualisation could be applied to trace continuous trade routes. These analytical approaches enable a more granular understanding of mobility patterns and facilitate predictions about unobserved areas based on inferred commodity flows.

While our vector field method offers significant advantages, certain limitations remain. The accuracy of interpolated vectors is influenced by the quality and density of the input data, with sparser regions showing higher uncertainty. In addition, this study focuses on a single commodity, leaving open the question of how well the method generalises to other trade systems or integrated datasets. Although our modelling is based on widely used origin–destination flow data, the approach is easily adaptable to other sources such as CDR and XDR records. These limitations point to promising directions for future research, including the use of higher-resolution data, the integration of multi-commodity flows, and applications to human mobility~\cite{barbosa2018human}. Addressing these challenges will further strengthen the versatility and impact of the vector field framework.

By transitioning from static network representations to dynamic vector fields, this research offers a versatile and scalable framework for studying commodity mobility. The approach not only advances our understanding of spatial flows but also establishes a foundation for future studies to refine and expand its applications in diverse fields, from public health to economic geography.

Critical points, such as sinks, sources, and saddle points, can provide deeper insights into the structural dynamics of commodity flows by identifying regions where trade activity converges, diverges, or transitions. Incorporating the identification of these points into the vector field framework enhances the understanding of how mobility patterns evolve over time and space, offering more detailed information about trade hubs, bottlenecks, and areas of influence. This analysis is particularly relevant for understanding how trade dynamics may impact broader systems, such as disease spread, resource allocation, or regional economic stability. The study of critical points also offers opportunities to integrate advanced vector field theories and analytical approaches for dynamic systems analysis. For instance, tracking the temporal evolution of these points could provide valuable insights into the stability and resilience of trade networks under changing conditions, such as seasonal variations or economic disruptions.

Our findings highlight a strong temporal variability in the spatial distribution of sink and source regions within the cattle trade system. Not only do the locations of these critical points shift across seasons, but their intensity also varies, as seen in the fluctuating density patterns in \figurename~\ref{fig:sink_source}. This temporal variation likely reflects the evolving roles of different premises and regions across seasons, influenced by market-driven dynamics and biological cycles such as breeding~\cite{paulino2014brazilian}. For instance, some premises may act as sources during active breeding or fattening periods, while others, such as slaughterhouses, consistently act as sinks due to receiving cattle without sending them onward. Notably, these critical points may sometimes emerge in regions not directly present in the observed origin–destination data, suggesting that depending on the commodity type or trade characteristics, influential areas may arise that are not explicitly recorded in the original dataset.

This seasonal variability could also reflect larger economic or logistical factors---such as transport infrastructure or policy changes---that influence the movement of cattle over time. Future research could explore how different modes of transportation or changes in trade regulations might influence the emergence, persistence, or reconfiguration of these sinks and sources throughout the year. Future work also could explore how the characteristics of critical points (\eg, their location, and movement) relate to the underlying drivers of trade, enabling targeted interventions or policy decisions.

\section*{Declaration}

For the purpose of open access, the authors have applied a Creative Commons Attribution (CC BY) license to any accepted manuscript version arising from this submission. 

\section*{Acknowledgment}

The authors acknowledge Dr. Denis Cardoso from Minas Gerais Institute of Agriculture (IMA) for providing the cattle trade dataset. ASM acknowledges the financial support from CNPq/Fapemig APQ-06591-24. 

\section*{Author Contribution Statement}

SF and RM developed the original ideas; All authors designed the study; RM supervised the development of the experiments; ASM wrote the formalisms; SF and MM collected, curated, and integrated the raw data; SF and MM performed the analysis; All authors analysed the results; SF and RM wrote the manuscript; SF and MM prepared the graphics; MM and ASM reviewed and revised the final version of the manuscript. All authors read, reviewed, and approved the final manuscript.

\bibliographystyle{plos2015}
\bibliography{bib.bib}

\begin{appendix}
\section*{Appendix}
\label{app:Appendix}

\setcounter{figure}{0}
\renewcommand{\figurename}{Fig.~A} 

\paragraph*{Dataset and code availability}
This study examines cattle movement in Minas Gerais, Brazil, using data from the Institute of Agriculture (IMA) covering 2013–2016. The dataset includes around 420,000 locations (e.g., farms, slaughterhouses, and markets) and records 3.78 million transactions for purposes like fattening, breeding, and trade. Each movement details the origin, destination, species, gender, age, and number of animals. Import and export data (less than 5\% of transactions) were excluded to focus on internal movement patterns.
For the analyses presented in this study, origin-destination (OD) matrices (both monthly and seasonal) were used. These, along with all the code for analysis and visualisation, are available at \url{https://github.com/Sima-Far/Flow_vector_field}.

\begin{figure}[!ht]
\centering
\includegraphics[width=.8\textwidth]{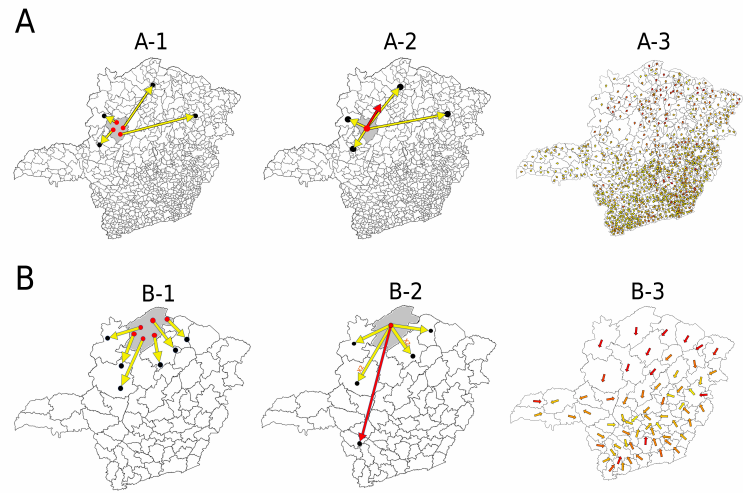}
\caption{\textbf{Generating vector fields from cattle trade data for (A) municipalities and (B) micro-regions.} \textbf{(A-1)} and \textbf{(B-1)} illustrate trade flows originating from a grey part (municipality or micro-region) to others. \textbf{(A-2)} and \textbf{(B-2)} represent the transformation of these trades into vectors, drawn from the centre of the grey area to the centres of destination areas, and then aggregated into a single resultant vector (shown in red). \textbf{(A-3)} and \textbf{(B-3)} demonstrate complete vector fields, where interpolation has been used to estimate vectors for municipalities or micro-regions lacking trade data during the selected time window. Different colours indicate varying vector magnitudes. This approach is applied for a specific time window and geographic division but can be adapted to different temporal or spatial granularities depending on analytical needs.}
\label{fig:vector_map_city_mic}
\end{figure}

\paragraph*{Interpolation of vector fields.}

There are multiple interpolation methods available for developing a continuous vector field. Among these, we opt for a triangle-based interpolation due to its computational efficiency and its suitability for handling the sparsity of our dataset's networks. Triangle-based interpolation enables us to estimate vector values at grid points where no initial vectors are present, while Delaunay triangulation efficiently manages sparse and irregularly spaced data by leveraging the geometric relationships between existing points~\cite{watson1984triangle}, ensuring a smooth and coherent representation of commodity flows.

\begin{figure}[ht]
\centering
\includegraphics[width=.6\textwidth]{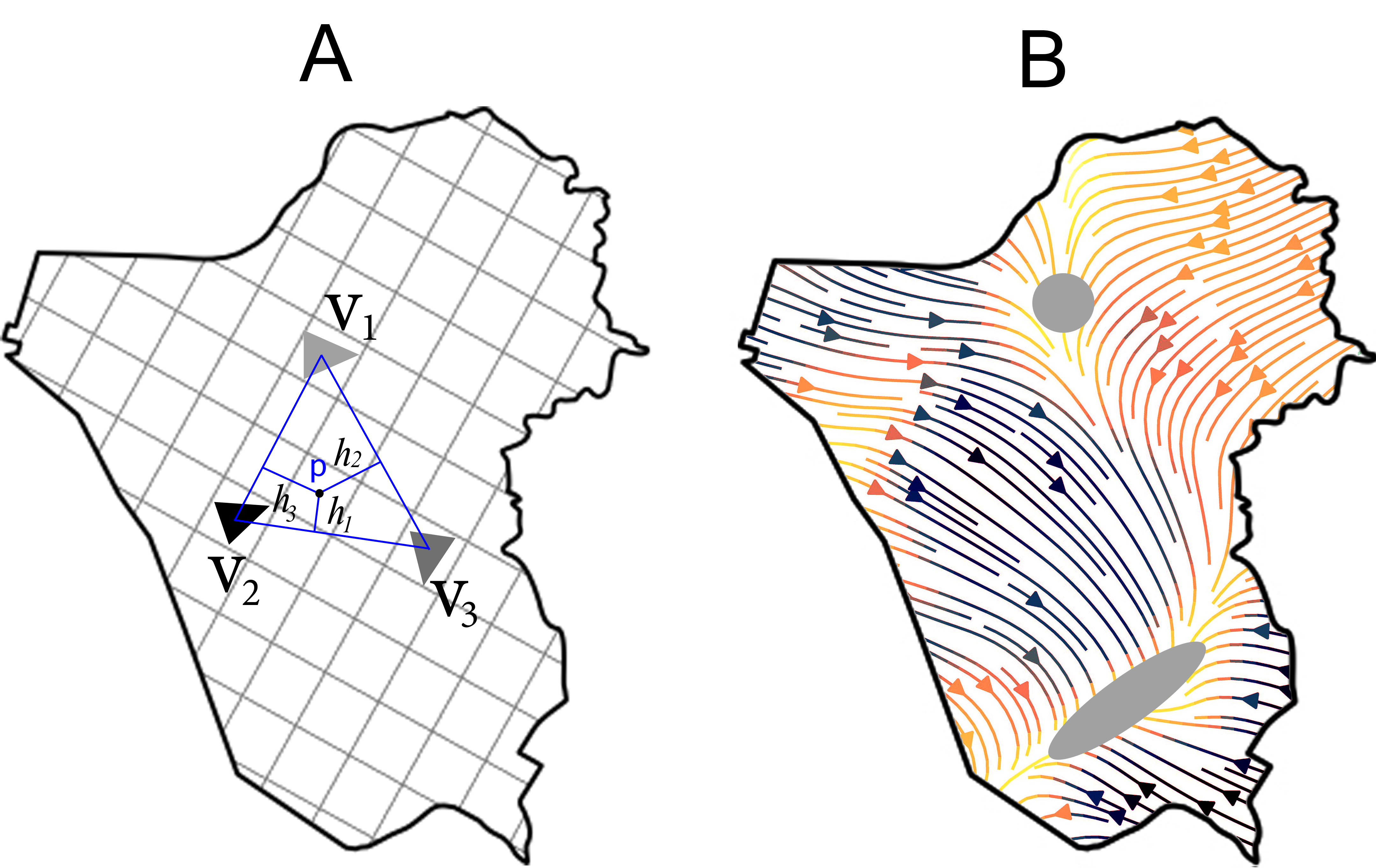}
\caption{\textbf{Critical points in the vector field.} 
{\bf (A)} Triangle-based interpolation method. To estimate vectors at specific locations within the triangulated mesh (generated using Delaunay triangulation on points with known vectors), we use a triangle-based interpolation technique. This method calculates the vector at a point $p$ by using the vectors at the three surrounding vertices.
{\bf (B)} Interpolated vector field. The resulting field displays interpolated vectors, with grey regions indicating critical points, which represent areas of attraction or repulsion. Colour variations reflect differences in vector magnitudes.}
\label{fig:critical_points_toy}
\end{figure}

To perform the interpolation, we use the centres of cells with initial vectors, along with selected boundary points assigned zero vectors to mitigate boundary effects. First, we triangulate the region using Delaunay triangulation on these points, ensuring that each point of interest lies within a well-defined triangular structure. Assume point $p$ falls inside a triangle with vertices having initial vectors $\mathbf{v}_1$, $\mathbf{v}_2$, and $\mathbf{v}_3$, as illustrated in \figurename~\ref{fig:critical_points_toy}A. If $h_1$, $h_2$, and $h_3$ represent the distances from point $p$ to the sides opposite to the angles at vertices with vectors $\mathbf{v}_1$, $\mathbf{v}_2$, and $\mathbf{v}_3$, respectively, the interpolated vector at point $p$, denoted as $\mathbf{v}_p$ is defined as~\cite{watson1984triangle,watson1985refinement}

\begin{equation}
\mathbf{v}_p=\frac{h_1}{h_1+h_2+h_3}\mathbf{v}_1 + \frac{h_2}{h_1+h_2+h_3}\mathbf{v}_2 + \frac{h_3}{h_1+h_2+h_3}\mathbf{v}_3\,.
\label{eqn:triangle}
\end{equation}

This interpolation method ensures that the resulting vector field is both continuous and adaptive to the underlying network structure. The final vector field representation, illustrated in \figurename~\ref{fig:critical_points_toy}B, captures both the main flows for each cell and the transitions across the entire field, offering a more detailed spatial understanding of commodity movements.

\paragraph*{Cosine similarity.} To quantify temporal patterns in flow directions, we use the cosine similarity measure which quantifies the directional consistency of commodity flows over time. For region $k$ between time intervals $i$ and $j$, the similarity is given by~\cite{beggs2012being}:
\begin{equation}\label{eq:CS}
S_C(\mathbf{v}^{i}_k, \mathbf{v}^{j}_k)=\frac{\mathbf{v}^{i}_k\cdot \mathbf{v}^{j}_k}{\|\mathbf{v}^{i}_k\|\|\mathbf{v}^{j}_k\|}\,,
\end{equation}
which represents the dynamic auto-correlation of the flow direction for region $k$. A value close to 1 indicates high directional stability over time, while lower values suggest greater variability.

The procedure consists of the following steps:
\begin{enumerate}
    \item Generating multiple vector fields over predetermined time intervals (\eg, months over four years).
    \item Computing cosine similarity values between consecutive intervals for each region.
    \item Constructing feature vectors for each region based on these similarity values.
    \item Applying clustering methods to group regions with similar temporal mobility patterns.
\end{enumerate}

This approach allows us to categorise regions based on the stability and evolution of their commodity flow directions, offering insights into the persistence and variability of trade movement patterns.

\paragraph*{Shannon entropy.} Shannon entropy is a measure of uncertainty or randomness associated with a set of possible outcomes. In a physical sense, it quantifies how dispersed or unpredictable a system is, akin to the concept of disorder in thermodynamics. It can be applied to assess the variability within a vector of values, capturing how evenly distributed different outcomes are~\cite{tsallis2009introduction}.

 Given a vector \( \mathbf{x} = [x_1, x_2, \ldots, x_n] \) where each \( x_i \) represents a category, event, or observed outcome, we first convert these values into a probability distribution. Let \( p_i \) denote the probability (or relative frequency) of outcome \( i \) occurring in the vector. The Shannon entropy \( H(\mathbf{x}) \) is then defined as~\cite{shannon1948mathematical}: 

\begin{equation}
H(\mathbf{x}) = -\frac{1}{\log n}\sum_{i=1}^{n} p_i \log p_i \,,
\label{eq:shannon_entropy}
\end{equation}
where the logarithm is typically taken in base 2 (yielding entropy in bits). This equation sums over all possible outcomes, weighting each by the negative logarithm of its probability. Higher entropy indicates greater diversity or unpredictability in the outcomes represented by the vector, meaning that commodity flow directions are more evenly distributed across different routes. Conversely, lower entropy suggests a more regular or predictable pattern, where flows concentrate in fewer, more predictable pathways. This process yields a single scalar value \( H(\mathbf{x}) \) that summarises the diversity of the vector's components.

\paragraph*{Critical points.} Critical points in a vector field are essential for understanding the field's structure and dynamics. These are points where the magnitude of the vector field becomes zero, i.e., ${\bf v}({\bf x})=0$, meaning all components of the vector vanish. In two-dimensional fields, these points are locations where field lines asymptotically converge or diverge, making the vector field's direction indeterminate.

A simple critical point is a special case where the vector field magnitude is zero at the point, but it does not vanish in its immediate surroundings. The behaviour near critical points can be studied using local linear approximations, assuming the vector field is smooth and differentiable. By employing a Taylor series expansion around a critical point ${\bf x}_0$, the local behaviour can be expressed as~\cite{smolik2016vector}:

\begin{equation}
    {\bf v}({\bf x}) = {\bf v}({\bf x}_{0}) + \frac{\partial {\bf v}}{\partial {\bf x}}({\bf x} - {\bf x}_{0})\,.
    \label{eq:approx}
\end{equation}

Because ${\bf x}_0$ is a critical point, ${\bf v}({\bf x}_{0})=0$. Equation~\ref{eq:approx} can then be rewritten in matrix form as follows:
\begin{equation}
\begin{bmatrix}
v_x\\
v_y
\end{bmatrix}
=
\begin{bmatrix}
\frac{\partial v_{x}}{\partial x}(x_0,y_0) & \frac{\partial v_{x}}{\partial y}(x_0,y_0)\\
\frac{\partial v_{y}}{\partial x}(x_0,y_0) & \frac{\partial v_{y}}{\partial y}(x_0,y_0)
\end{bmatrix}
\begin{bmatrix}
x-x_0\\
y-y_0
\end{bmatrix}\,,
\label{eq:approx_matrix}
\end{equation}
\begin{equation}
    {\bf v} = {\bf J} \cdot ({\bf x}-{\bf x}_{0})\,.
    \label{eq:jacobi_v}
\end{equation}

The Jacobian matrix, ${\bf J}$, is a mathematical representation that captures how a vector field changes near a critical point. It plays a key role in classifying critical points by analysing their local behaviour. By studying the eigenvalues and eigenvectors of ${\bf J}$, we can understand how tangent curves behave around the critical point.

\paragraph*{Moran's $I$.}  
The Moran's $I$ statistic quantifies the degree of spatial autocorrelation by analysing both the values and locations of observations. It assesses whether a variable exhibits spatial clustering, dispersion, or randomness, summarising spatial patterns within a dataset. The Global Moran's \(I\) index is calculated as follows~\cite{rey2023geographic}:
 \begin{equation}
 I= \frac{n}{\sum_{i}\sum_{j} w_{ij}} \times \frac{\sum_{i}\sum_{j}w_{ij}z_{i}z_{j}}{\sum_{i}z_{i}^2}\,,
 \label{eq:moran}
 \end{equation}

where:
\begin{itemize}
    \item \( n \) is the number of observations,
    \item $z_{i}$ is the standardised value of the observation at location $i$ (commonly $z_{i}= y_{i} - \bar{y}$),
    \item \( y_i \) and \( y_j \) are values at locations \( i \) and \( j \), respectively,
    \item \( \bar{y} \) is the mean of all values,
    \item \( w_{ij} \) represents the spatial weights matrix, defining the spatial relationship between pairs of locations.
\end{itemize}

Moran’s \(I\) values range from -1 to +1:
\begin{itemize}
    \item \textbf{Positive values} indicate spatial clustering, where similar values are near each other.
    \item \textbf{Negative values} suggest spatial dispersion, where high values are situated near low values, and vice versa.
    \item \textbf{Values close to zero} imply randomness, with no discernible spatial pattern in the data.
\end{itemize}

As a graphical tool, the Moran Plot provides a valuable resource for understanding both global and local spatial autocorrelation. These plots take the form of scatter plots, where values of interest are plotted against spatial lag (as shown in \figurename~\ref{fig:Moran_I}A) for each location~\cite{yohman2021}.

Spatial lag variable represents weighted sums or averages of neighbouring values. The spatial lag characterises how a variable behaves near each location, essentially acting as a local smoother. This can be expressed in matrix notation as follows~\cite{anselin1995local, rey2023geographic, darmofal2015spatial}:
\begin{equation}
    y_{sl-i}= \sum_{j}w_{ij}y_{j} \,.
    \label{eq:slag_matrix}
\end{equation}

In Equation~\ref{eq:slag_matrix}, $y_{sl-i}$ represents the spatial lag for location $i$, calculated as the weighted sum of values at all other locations. As mentioned before, $y_j$ is the value in location $j$, and $w_{ij}$ is the entry in the $i$-th row and $j$-th column of a $W$ spatial weights matrix. Since non-neighbours receive a weight of zero, $y_{sl-i}$ effectively captures the weighted average of values at observation $i$'s neighbours. Spatial lag of the vector magnitudes is calculated using spatial weights based on Queen Weights~\cite{berry1968spatial}. This weight reflects adjacency relationships using a binary variable that indicates whether a polygon shares an edge or a vertex with another polygon. The matrix expression for this weight is as follows:

\begin{equation}
    W_{ij}= \left\{ \begin {array} {cc} 1, & \text{if regions} \ i \ \text{and} \ j \ \text{share an edge or vertex,}  \\ 0, & \text{otherwise.} \end{array} \right.\,
\end{equation}

\end{appendix}

\end{document}